\documentclass[twocolumn,letter]{jpsj2} 

\def\nle{\ \raise.3ex\hbox{$<$}\kern-0.8em\lower.7ex\hbox{$\sim$}\ }
\def\nge{\ \raise.3ex\hbox{$>$}\kern-0.8em\lower.7ex\hbox{$\sim$}\ }
\def\chitwo{\chi''(\omega; t)}
\def\hcr{h_{\rm cr}}
\def\lcrh{L_h}
\def\lomega{L_{\omega}}
\def\Rcr{R_{\rm cr}}

\def\Rtr{R_{\rm tr}}
\def\Rw{R_{\rm w}}
\def\Tc{T_{\rm c}}
\def\Tcr{T_{\rm cr}}
\def\tcr{t_{\rm cr}}
\def\Tf{T_{\rm f}}
 
\def\tw{t_{\rm w}} 
\def\twcr{t_{\rm w}^{\rm cr}} 
\def\fcm{M_{\rm FC}}

\def\zfcm{M_{\rm ZFC}}
\def\FeMnTiO{Fe$_{0.5}$Mn$_{0.5}$TiO$_3$}

\title{Field-Shift Aging Protocol on the 3D Ising Spin-Glass Model:\\
Dynamical Crossover between the Spin-Glass and Paramagnetic States}
\author{ Hajime \textsc{Takayama}$^{1}$\footnote{E-mail:takayama@issp.u-tokyo.ac.jp} 
and Koji \textsc{Hukushima}$^{2}$\footnote{E-mail: hukusima@phys.c.u-tokyo.ac.jp} }
\inst{$^{1}$Institute for Solid State Physics, University of Tokyo,
5-1-5 Kashiwa-no-ha, Kashiwa, Chiba 277-8581 \\
$^{2}$ Department of Basic Science, Graduate School of Arts and
Sciences, University of Tokyo, 3-8-1 Komaba, Tokyo 153-8902}
\recdate{\today}
\abst{
Spin-glass (SG) states of the 3-dimensional Ising Edwards-Anderson 
model under a static magnetic field $h$ are examined by means of the 
standard Monte Carlo simulation on the field-shift aging protocol at
temperature $T$. For each process with $(T; \tw, h)$, $\tw$ being the
waiting time before the field is switched on, we extract the dynamical
crossover time, $\tcr(T; \tw, h)$. We have found a nice scaling 
relation between the two characteristic length scales which are properly
determined 
from $\tcr$ and $\tw$ and then are normalized by the static field 
crossover length introduced in the SG droplet theory. This scaling 
behavior implies the instability of the SG phase in the equilibrium 
limit even under an infinitesimal $h$. In comparison with this 
numerical result the field effect on real spin glasses is also 
discussed.  
}
\kword{spin glass, aging phenomena, droplet picture, Ising EA model, Monte Carlo simulation}

\begin{document}
\maketitle

In spite of extensive studies in more than a decade the nature of 
ordering in spin glasses have remained unsettled~\cite{review1}. A most 
notable problem is on the stability of the equilibrium spin-glass (SG)
phase in a static field $h$. The mean-field theory predicts its
stability up to a certain critical magnitude of $h$~\cite{AT-line},
while by the droplet theory~\cite{FH-88-NE,FH-88-EQ,BM-chaos} the SG
phase is unstable even in an infinitesimal $h$. In our opinion, a
fundamental difficulty of the SG studies lies on the following fact: spin 
dynamics governed by thermally activated processes is so slow that the 
time window of experiments on real spin glasses, not to mention that of
numerical simulations, is insufficient for us to distinguish whether a
SG-state we are observing is in an asymptotic regime close to equilibrium 
or far from it. In such circumstances, a possible strategy is
a literal numerical experiment on nonequilibrium dynamics of model 
spin glasses without referring to their equilibrium properties which have 
been extensively looked for by some ingenious, but artificial 
methods~\cite{Hukushima,genetic,KHMMP}. Following this strategy, we have 
been extensively studying aging phenomena in the three-dimensional (3D) 
Ising Edwards-Anderson (EA) model~\cite{ours1,ours2,oursSupp,ours3}
including the comparison of the results with those measured in real spin
glasses~\cite{oursLoren}. 

In the present work we address the problem of the SG-phase stability in
a field by the strategy mentioned above, concentrating on the following 
field-shift ($h$-shift) aging protocol. A system is instantaneously
quenched at $t=0$ from $T=\infty$ to a temperature $T$ below the
transition temperature $\Tc$, it is aged in zero field until $t=\tw$
when $h$ is switched on, and the induced magnetization, called 
zero-field-cooled (ZFC) magnetization and denoted as $\zfcm(t')$, is
measured as a function of $t'=t-\tw$. At $t < \tw$ isothermal and
isobaric aging under $h=0$ proceeds and the SG short-range order, or the
mean size of SG domains, $R_T(t)$, grows
as~\cite{ours1,Kisker-96,Marinari-growthLaw},  
\begin{equation}
R_T(t) = b_Tt^{1/z(T)},
\label{eq:RTt}
\end{equation}
with $b_T$ and $z(T)$ being constants. The characteristic time scale 
after the $h$-shift most frequently examined is the one where the 
logarithmic time derivative of $\zfcm(t')$, i.e., 
$S(t') = \partial (\zfcm(t')/h) /\partial {\rm ln}t'$ exhibits a peak. 
This behavior of $S(t')$ is interpreted to represent a crossover from
the isobaric aging state under $h=0$ to that under $h>0$. The peak
position of $S(t')$ is called here the crossover time and denoted as
$\tcr$, though it has been frequently called the effective waiting  
time. It is a function of $T,\ \tw$ and $h$. We then expect that in the  time
range $t' \nle \tcr$, which we call the {\it transient regime} of $h$-shift
processes, SG subdomains in local equilibrium with respect to $h>0$ grow 
within each SG domains which were in local equilibrium with respect to
$h=0$ at $t=\tw$. Furthermore we {\it intuitively} introduce the growth
law of mean size of the subdomains in the transient regime as
\begin{equation}
\Rtr(t') = b_Tt'^{1/z(T)-a_Th^2}\ \ \ \ \ (t' \nle \tcr).
\label{eq:Rtr}
\end{equation}
Compared with the growth law of Eq.(\ref{eq:RTt}), the second term with
$a_T$ is added in the exponent. 

The last step of our analysis is to relate the two characteristic length 
scales above-introduced, $\Rw \equiv R_T(t=\tw)$ and 
$\Rcr \equiv \Rtr(t'=\tcr)$, to the {\it field crossover length}, 
$\lcrh$ introduced in the droplet theory~\cite{FH-88-EQ}. The latter
separates characteristic behavior of droplet excitations in equilibrium
by their size $L$, such that it is dominated by the Zeeman energy 
($\sim hL^{d/2}$) for $L>\lcrh$ and by the SG free energy gap 
($\sim \Upsilon L^\theta$) for $L<\lcrh$. Here $d$ is the spatial 
dimension, $\Upsilon$ the stiffness constant of domain walls, and
$\theta$ the gap exponent. Explicitly $\lcrh$ is written as  
\begin{equation}
\lcrh = l_T h^{-\delta}, 
\label{eq:lcrh}
\end{equation}
where $\delta = (d/2 - \theta)^{-1}$ and $l_T$ is a constant. 
Surprisingly it is found that, with a proper choice of the values of
$a_T$, the scaled plot $\Rcr/\lcrh$ vs $\Rw/\lcrh$ of all the data
obtained for different $T,\ \tw$ and $h$ lie on a single curve. The
consequence of this scaling is clear; the $h$-shift aging process is
nothing but a dynamical crossover from the SG state in $h=0$ to the
paramagnetic state in $h>0$, or, the SG phase is unstable under $h>0$ in 
the equilibrium limit. Furthermore, when the simulated results are
simply extended to the time window of the experiment on a real Ising
spin glass \FeMnTiO\, our picture of the dynamical crossover turns out
to be consistent with the experimental data which were interpreted as an
evidence of the AT phase transition predicted by the mean-field
theory~\cite{KatoriIto94}. 

We carry out standard heat-bath Monte Carlo (MC) simulations on the 3D 
Ising EA model with interactions obeying a Gaussian distribution with
mean zero and variance $J$ which plays a role of units of energy and 
temperature (with $k_{\rm B}=1$). The strength of field $h$ is 
represented by its associated Zeeman energy. The time is measured in
unit of the MC steps per spin (mcs). In the present work we use the 
values of parameters $b_T$ and $z(T)$ in Eq.(\ref{eq:RTt}) obtained 
previously~\cite{ours1}; $(T; b_T, z(T))=$ (0.4; 0.82, 14.8), 
(0.5; 0.80, 11.8), (0.6; 0.78, 9.8), (0.7; 0.78, 8.7) and (0.8; 0.76,
7.9). Note that $\Tc \simeq 0.95$~\cite{Marinari-cd98-PS,MariCamp}
for this SG model. In the temperature range $T \le 0.8$ and in the time
range $t \nle 10^5$ where we carry out simulations, $R_T(t)$ is less
than a few lattice distances~\cite{ours1}. Therefore we examine systems
with a linear size of 24, but an average over at least 6400 samples with 
different sets of interactions is taken for evaluating physical
quantities in each $h$-shift process discussed below. 

\begin{figure}[b]
\begin{center}
\resizebox{0.5\textwidth}{!}{\includegraphics{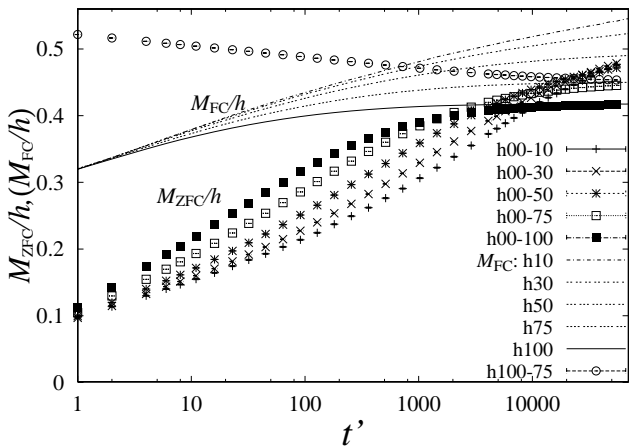}}
\end{center}
\caption{ $\fcm(t)$ ($t'=t$), $\zfcm(t')$ with $\tw=4096$, and $M(t')$
 under $h=0.75$ after the shift from $h=1.0$ ($\tw=4096$).}
\label{fig:Ms-t4096}
\end{figure}

Let us begin our discussions with the typical results of $\zfcm(t')$
and $\fcm(t')$, the field-cooled (FC) magnetization defined by
$\zfcm(t')$ with $\tw=0$, at $T=0.6$ shown in
Fig.~\ref{fig:Ms-t4096}. Both exhibit a jump at $t'=1$, which is
naturally attributed to one-spin flips of individual spins whose
internal field at $t'=0$ is smaller than and opposite to $h$. The number
of such spins decreases in the ZFC process at $t \le \tw$ and so 
$\fcm(1) > \zfcm(1)$. After the jump the system ages to an isobaric
aging state under $h$ gradually and irreversibly. Associated with this,
subdomains in local equilibrium with respect to $h$ are expected to grow
obeying Eq.(\ref{eq:Rtr}), though we have not succeeded to observe them
directly in simulation yet. Also shown in the figure is $M(t')$ which is
observed in $h=0.75$ after shift-down from the FC process with $h=1.0$
of a period $\tw$. One can see that the magnetizations under 
$h=0.75\ (1.0)$ at $t'\nge 10^6\ (10^4)$ are independent of the history
of the system at $t' \le 0$, i.e., the system reaches the equilibrium
paramagnetic state at this time range. For a small $h$ such as $h=0.1$,
on the other hand, we cannot see even a precursor of saturation in
$\fcm(t')$. Our interest is therefore in behavior of $\zfcm(t')$ for 
$0.1 \nle h \nle 0.75$.

\begin{figure}[t]
\begin{center}
\resizebox{0.5\textwidth}{!}{\includegraphics{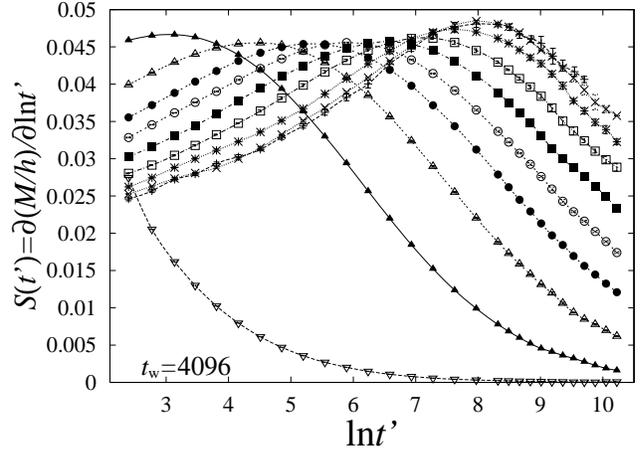}}
\end{center}
\caption{ $S(t')$ for the $h$-shift processes at $T=0.6$ with
 $\tw=4096$ plotted vs  ln$t'$. The data are for
 $h=$0.05, 0.1, 0.2, 0.3, 0.4, 0.5, 0.6, 0.75, 1.0 and 2.0 from right to
 left.
 }
\label{fig:Soft-up}
\end{figure}

In Fig.~\ref{fig:Soft-up} we demonstrate a set of $S(t')$ observed in the 
$h$-shift processes at $T=0.6$ for various $h$ but with a fixed $\tw$. 
Except for $h=2.0$, $S(t')$ exhibits a peak at $t'=\tcr$. All $\tcr$
obtained for various sets of $h$ and $\tw$ are shown in the inset of
Fig.~\ref{fig:Rcr-scal-up} below. For smaller fields, $h \nle 0.1$, we
obtain $\tcr \simeq \tw$, as has been observed since the very beginning
of the SG aging study~\cite{Lundgren}. For larger $h$, however, $\tcr$
becomes significantly smaller than $\tw$.

As noted at the beginning of this letter, a key quantity in the droplet 
theory on the SG aging in a field is the field crossover length,
$\lcrh$, and its role on the dynamical crossover of present interest
becomes apparent if we compare it with the characteristic length scales, 
$\Rcr$ and $\Rw$, of each $h$-shift process. 
As already defined, the latter are a function of $T,\ 
\tw$ and $h$. In processes with such a large $h$ or/and $\tw$ that 
$\Rw \gg \lcrh$ holds, the aging dynamics after the $h$-shift is
dominated by the Zeeman energy and so $\Rcr \simeq \lcrh$ is expected
irrespectively of $\tw$. In opposite processes with $\Rw \ll \lcrh$, we 
expect $\Rcr \simeq \Rw$. These qualitative features among $\Rw, \Rcr$, 
and $\lcrh$ are made more transparent by the scaling plots of 
$\Rcr/\lcrh$ vs $\Rw/\lcrh$. Even if we simply put 
$\Rtr(t')=R_T(t')$~\cite{mine}, i.e., $a_T=0$ in Eq.(\ref{eq:Rtr}), we
obtain the scaling plot, from which we can see the tendency of
saturation of $\Rcr/\lcrh$ at large $\Rw/\lcrh$.   

\begin{figure}[t]
\begin{center}
\resizebox{0.5\textwidth}{!}{\includegraphics{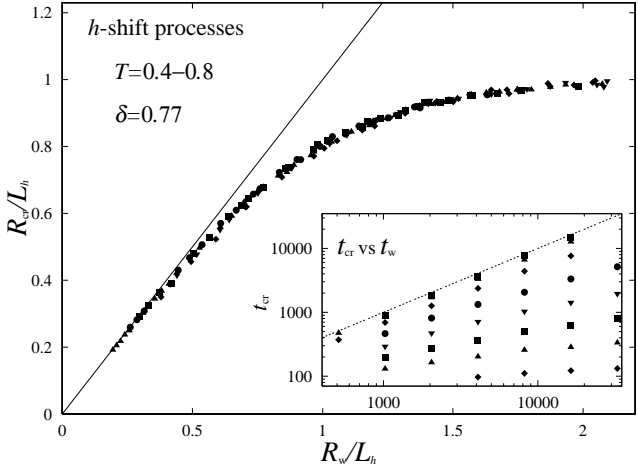}}
\end{center}
\caption{Scaling plots of $\Rcr/\lcrh$ vs $\Rw/\lcrh$ of the $h$-shift 
processes at $T=0.4 \sim 0.8$ with $a_T$ properly chosen. 
The line represents $y=x$. In the  inset we plot $\tcr$ vs. $\tw$ in the
 $h$-shift processes at $T=0.6$  
with $h=$0.05, 0.1, 0.2, 0.3, 0.4, 0.5, 0.6, and 0.75 from top to bottom. 
}
\label{fig:Rcr-scal-up}
\end{figure}

More interestingly, as shown in Fig.~\ref{fig:Rcr-scal-up}, all the 
sets of ($\Rcr/\lcrh, \Rw/\lcrh$) obtained in the $h$-shift processes at 
different temperatures, $T=0.4, 0.5, 0.6, 0.7,$ and 0.8, with
different sets of $\tw$ and $h$, lie on a single 
curve, if the values of $a_T$ and $l_T$ are properly adjusted as
$(T; a_T, l_T)=$ (0.4; 0.06, 0.85), (0.5; 0.075, 0.83), 
(0.6; 0.09, 0.81), (0.7; 0.115, 0.78) and (0.8; 0.125, 0.75). We
emphasize here that the values of the parameters other than $a_T$ and
$l_T$ are those previously determined independently of the present
analysis; $\theta = 0.20$,~\cite{ours1,BM-theta} and so 
$\delta \simeq 0.77$ in Eq.~(\ref{eq:lcrh}) with $d=3$, and 
$(b_T, 1/z(T))$ already listed above. 
The adjustment of $a_T$ is essential to obtain a unique 
scaling curve for the data set at each temperature. The proper choice of 
$l_T$, which commonly scales both axes at each temperature, makes the
scaling curves at different temperatures to lie top on each
other. Physically, the main $T$-dependence of $l_T$ can be attributed to
that of the stiffness constant $\Upsilon$ which is expected to vanish at
$\Tc$. 

The scaling behavior we have found implies that, for $h$ of any
strength, if  a system is aged under $h=0$ up to $\tw$ for which 
$\Rw \sim \lcrh$ is satisfied, the aging dynamics after the $h$-shift is
governed dominantly by the Zeeman energy, i.e., the system exhibits
paramagnetic behavior at $t' \nge \tcr$. If $\tw$ is smaller than such a
value, say $\tw=0$, it is faster for the system to reach the
paramagnetic state as demonstrated in Fig.~\ref{fig:Ms-t4096}.

To reach the nice scaling shown in Fig.~\ref{fig:Rcr-scal-up}, the term 
intuitively added in the exponent in Eq.(\ref{eq:Rtr}) plays an essential 
role, though the scaling behavior is not so sensitive on the precise 
value of $a_T$, i.e., even if it is changed by $\pm 0.01$ from the
value listed above, the quality of the scaling does not become 
significantly worse. A positive $a_T$ implies that the subdomains in the
transient regime grow more slowly than the domains do in the isobaric
aging, i.e., $\Rtr(t=t') < R_T(t)$. We consider it quite plausible that 
the domain growth in $h$ starting from a state already aged in $h'\ne h$ 
(including $h'=0$) is slower than that starting from a state not aged at 
all. In this context it is worth noting that Joh {\it et al.} explained 
their experimental result $\tcr < \tw$ for relatively large $h$ in terms 
of the reduction of the barrier energy for excitations due to the Zeeman 
energy~\cite{Joh99}. Numerically we have also observed that $\tcr < \tw$, 
but our interpretation for it is quite different from theirs.

Our MC simulation alone cannot exclude a possible existence of the SG 
phase under sufficiently small fields $(h \nle 0.1)$. However, the 
deviation of 
$\Rcr$ from $\Rw$ is clearly recognized even at small 
$x \equiv \Rw/\lcrh$. It originates from the fact that the excitation 
gaps of droplets with a fixed size distribute continuously and a finite 
weight exists even at zero energy. This point has been recently 
emphasized by J\"onsson {\it et al}~\cite{PYN-02-R} who have called the 
consequence of these excitations the {\it weak-chaos effect}. Our data 
of $\Rcr/\lcrh$ vs. $\Rw/\lcrh$ at small, though not sufficiently small, 
$x$ are consistent with their argument. We may better call this the 
{\it weak-paramagnetic effect}.

\begin{figure}
\begin{center}
\resizebox{0.5\textwidth}{!}{\includegraphics{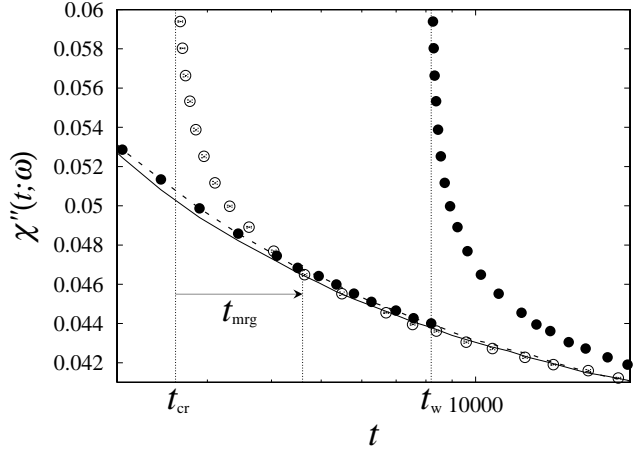}}
\end{center}
\caption{$\chitwo$ at $T=0.6$ for $\tau_\omega=2\pi/\omega=64$, $\tw=8192$, and
 $h=0.3$ (filled circles). When its branch at $t>\tw$ is moved to the
 left by the amount  $t_{\rm mv}=\tw-\tcr=5600$ (open circles) it
 merges to the reference curve of $h=0.3$ (solid line) at 
$t'= t_{\rm mrg} \simeq 2000 \sim \tcr$ (see Ref. 12 for further details of
 determining $t_{\rm mrg}$). The dashed line represents the 
 reference curve of $h=0$. Those for $h=0.05, 0.1, 0.2$ (not shown)
 almost coincide with the reference curve of $h=0$.}
\label{fig:Sus-up}
\end{figure}

One more comment is in order on the out-of phase component of the ac 
susceptibilities, $\chitwo$, measured in the $h$-shift aging protocol. It 
is evaluated via the fluctuation-dissipation theorem~\cite{ours2}, and 
its typical behavior is shown in Fig.~\ref{fig:Sus-up}. In the isobaric 
regime at $t \le \tw$, $\chitwo$ is rather insensitive to $h$ as seen in 
the figure. Just after the $h$-shift, however, it exhibits a sharp jump-up, 
and subsequently it tends to merge to the one observed in the isobaric 
aging under $h$ from $t=0$, which we call the reference curve. 
Qualitatively similar behavior has been observed  
experimentally~\cite{h-Eric95}. According to the droplet theory, the aging 
part of $\chitwo$ in the isobaric regime is given by 
a function of $\lomega/R_T(t)$, where $\lomega$ is a mean size of
droplets which can respond to the ac field of frequency 
$\omega$~\cite{FH-88-NE,ours2}. A naive extension of this idea 
is that $\chitwo$ after the $h$-shift is dominantly governed by $\Rtr(t')$, 
the mean size of subdomains which is smaller than $\Rw$. In fact, as 
demonstrated in Fig.~\ref{fig:Sus-up}, the time scale that the $\chitwo$ 
merges to the reference curve well coincides with $\tcr$, the peak position 
of $S(t')$. Thus the following picture on the aging dynamics around 
$t' \sim \tcr$ is extracted. At $t' \nle \tcr$ subdomains of mean size 
$\Rtr(t')$ grow up obeying Eq.(\ref{eq:Rtr}), and at $t'\nge \tcr$ the
subdomains become main domains in the system. For a process with 
$\Rcr \simeq \lcrh$ the system is regarded to be already in a
paramagnetic state and SG domains do not grow any more. For a process
with $\Rcr < \lcrh$, on the other hand, the mean size of SG domains 
grows following
Eq.(\ref{eq:RTt}) with a shift of the time origin by a certain amount. 
This growth is observed through physical quantities, such as $\chitwo$
in Fig.~\ref{fig:Sus-up}, which associate a length
scale smaller than the mean size of domains at $t'$. The value of 
$\zfcm(t')$ at $t' \sim \tcr$ is, however, still significantly
smaller than $\fcm(t')$, and it needs further long time for $\zfcm(t')$
to catch up $\fcm(t')$, or the mean size of SG domains reaches $\lcrh$.  

So far described are the results of our simulation on the EA Ising SG
model. In order to look for their relevance to phenomena in real Ising 
spin glasses, let us introduce the dynamical crossover condition 
defined by $\Rw(\tw^{\rm cr})/\lcrh=x_{\rm c}$ with $x_{\rm c}$ being
a constant nearly equal to unity. It is rewritten as   
\begin{equation}
\hcr = c_{\Tcr} (\twcr/t_0)^{-1/\delta z(\Tcr)},
\label{eq:length-h}
\end{equation}
with $c_T=(x_{\rm c}l_T/b_T)^{1/\delta}$ and $t_0$ being the microscopic
time unit. This equation gives us an estimate of the value $\hcr,\ \Tcr$
or $\twcr$ when the values of the other two parameters are specified,
and when $h \nge(\nle) \hcr$, for example, the system is regarded as in
a paramagnetic (SG) state. 

\begin{figure}
\begin{center}
\resizebox{0.4\textwidth}{!}{\includegraphics{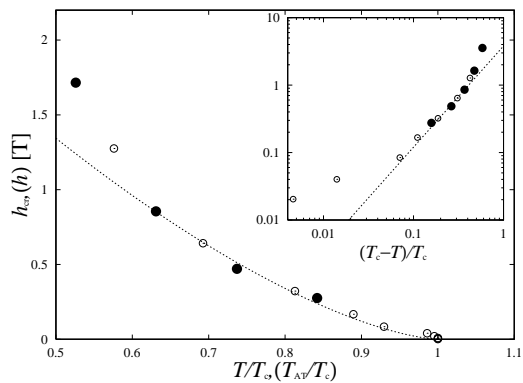}}
\end{center}
\caption{Dynamical crossover field $\hcr$ (solid circles) due to Eq.(4)
 and the AT critical points (open circles) determined experimentally in
 [16] for \FeMnTiO\ under fields. Shown in the inset are the same data
 plotted w.r.t. $(\Tc-T)/\Tc$ double-logarithmically. The line
 represents the AT critical line.}
\label{fig:h-cr-T}
\end{figure}

Katori and Ito (KI)~\cite{KatoriIto94} determined the critical temperature, 
$T_{\rm AT}$, of Ising spin glass \FeMnTiO\ under $h$ as
the one where the $\zfcm$ starts to deviate from $\fcm$ within the 
observation time of $t_{\rm ob} \simeq 10^2$~s at each temperature. Regarding 
this as the dynamical crossover with $\twcr=t_{\rm ob}$ and 
$T_{\rm AT}=\Tcr\ (=T)$, we evaluate $\hcr$ by Eq.(\ref{eq:length-h}) 
with $\Tc=21.5$~K and $t_0=10^{-12}$~s. This means that we simply extend the 
result extracted from our simulation to the time range of $10^{14}$~mcs as 
we did with success in our previous work~\cite{oursLoren}. Surprisingly, as 
shown in Fig.~\ref{fig:h-cr-T}, the $\hcr - T$ plot thus evaluated with 
$x_{\rm c} \simeq 1.5$ coincides very well with the $h - T_{\rm AT}$
plot that KI  
regard the AT critical line~\cite{AT-line}.  Similarly, the freezing 
temperatures $\Tf(\omega,h)$ extracted by Mattsson 
{\it et al.}~\cite{Mattsson95} through the ac susceptibility measurement 
are also satisfactorily reproduced by our analysis with $\Tcr=\Tf,\  
\twcr=2\pi/\omega$ and $\hcr=h$. These agreements, whose details will be 
reported elsewhere, are quite promising for us to consider that the field 
effect also on real spin glasses is well described by our dynamical 
crossover picture~\cite{mine}.  For $h$ of the magnitude less than 1 Oe,
however, $\twcr$ becomes astronomically large, which explains, we
consider, why $\zfcm/h$ measured in laboratory-time deviate from
$\fcm/h$, and $\fcm/h$ itself does so from the paramagnetic
susceptibility at $T\simeq \Tc$ for small $h$~\cite{Nagata}.  

To conclude we have demonstrated that the characteristic length scales 
associated with $h$-shift aging protocols in the 3D EA Ising SG model
exhibit unique scaling behavior. Because of the latter, we have
succeeded to reach one of the most important equilibrium properties of
Ising spin glasses via simulations on nonequilibrium 
(aging) dynamics: their SG phase is unstable under a finite field in
equilibrium.  

We thank to P. J\"onsson and H. Yoshino for fruitful discussions. The 
present work is supported by a Grant-in-Aid for Scientific Research
Programs (\# 14540351 and \#14084204) and NAREGI Nanoscience Project,
all from the 
Ministry  of Education, Science, Sports,
Culture and Technology. The numerical simulations have been
performed using the facilities at the Supercomputer Center, Institute
for Solid State Physics, the University of Tokyo.

\end{document}